\documentstyle[12pt,epsfig]{article}

\newfont{\Bbb}{msbm10 scaled 1200}     

\newcommand{\mysection}[1]{\section[#1]{#1}}

\renewcommand{\abstract}[1]{\vspace*{1in}
\centerline{\bfseries\normalsize\abstractname}\bigskip
{\renewcommand{\baselinestretch}{1.2}\normalsize #1} \newpage}

\newcommand{\ben}{\begin{enumerate}}
\newcommand{\een}{\end{enumerate}}
\newcommand{\eq}[1]{\begin{equation} #1 \end{equation}}

\begin{document}

\begin{center}

{\Large \bf Fluctuations of Strangeness \\
            and \\
            Deconfinement  Phase Transition \\
\vspace{0.2cm}
            in Nucleus--Nucleus Collisions}

\vspace{1.5cm}


{\bf {\bf M.I. Gorenstein}$^{a,b}$, M. Ga\'zdzicki}$^{c,d}$ and {\bf O.S.
Zozulya}$^{b,e}$

\vspace{1cm}

\noindent
\begin{minipage}[t]{12.5cm}
$^{a}$ Bogolyubov Institute for Theoretical Physics, Kiev, Ukraine\\
$^{b}$ Institut f\"ur Theoretische Physik, Universit\"at Frankfurt,
Germany\\
$^c$ Institut f\"ur Kernphysik, Universit\"at Frankfurt, Germany\\
$^d$ \'Swi\c{e}tokrzyska Academy, Kielce, Poland \\
$^e$ Taras
Shevchenko Kiev National University, Kiev, Ukraine\\
\end{minipage}
\end{center}


\abstract{
We suggest that the
fluctuations of strange hadron multiplicity could be sensitive to the
equation of state and microscopic structure of strongly interacting matter
created at the early stage of high energy nucleus--nucleus collisions.
They may serve as an important tool in the study of the deconfinement
phase transition. We predict, within the statistical model of the early
stage, that the ratio of properly filtered 
fluctuations of strange to non-strange hadron
multiplicities should have a non--monotonic energy dependence with a
minimum in the mixed phase region.

}

\newpage

\mysection{Introduction}

Recently a new method was proposed \cite{Gazdzicki:2003bb} to study the
equation of state (EoS) of the matter created at the early stage of
nucleus--nucleus (A+A) collisions. It was suggested to analyze the
collision energy dependence of the ratio of properly filtered multiplicity
and energy fluctuations. It was shown that the fluctuation ratio measured
at the final state may be directly dependent on the ratio of pressure to
energy density at the early stage. In the present paper we make a next
step within the same framework and consider strangeness fluctuations. A
new observable connected with the fluctuations of the strange hadron yield
is proposed. We derive the quantitative predictions of the Statistical
Model of the Early Stage (SMES) \cite{Gazdzicki:1998vd} concerning the
fluctuation of the number of strange hadrons. We find a non-monotonic
dependence on the collision energy: the ratio of the relative multiplicity
fluctuations of strange hadrons to those of negatively charged hadrons has
a {\it minimum} in the domain where the onset of deconfinement occurs.

The paper is organized as follows. The basic relevant assumptions of the
SMES are presented in Sect. 2. The role of dynamical fluctuations in the
study of EoS at the early stage is stressed in Sect. 3. The energy
dependence of the dynamical strangeness fluctuations is derived within
SMES and discussed in Sect. 4. Summary and conclusions close the paper,
Sect. 5.

\mysection{Statistical Model of the Early Stage}

Since we are going to discuss the collision energy dependence of the
fluctuations within the SMES \cite{Gazdzicki:1998vd}, let us present its
basic assumptions. The volume, $V$, where the matter in confined, mixed or
deconfined state is produced at the collision early stage, is given by the
Lorentz contracted volume occupied by wounded nucleons. For the most
central collisions the number of wounded nucleons is $N_W \approx 2 \cdot
A$. The net baryon number of the {\em created} matter equals to zero. Even
in the most central A+A collisions only a fraction $\eta <1$ of the total
collision energy is used for a particle production. The rest is carried
away by the baryons which contribute to the baryon net number. The main
postulates of the SMES \cite{Gazdzicki:1998vd} are the following:
\begin{itemize}
\item New particles are created at the early stage of A+A collision
in a state of the global statistical equilibrium.
\item The model assumes that the {\it created} matter is described
in the grand canonical ensemble with all chemical potentials  equal
to zero. The EoS is chosen in the form of relativistic ideal gas with an
additional bag--term contribution in the deconfined phase.
\item The basic constituents of the deconfined phase are the
light $u$ and $d$ (anti)quarks ($m_{u}\cong m_{d}\cong 0$), the strange
(anti)quarks ($m_{s}\cong 175$~MeV) and gluons.
\item The total entropy and strangeness created
at the early stage are supposed to be approximately conserved
 during the expansion, hadronization and freeze-out.
\end{itemize}

We describe the system's EoS in terms of the pressure as a function of
temperature: $p = p(T)$. For an ideal gas of particles `$j$' with zero
chemical potential one finds ($m_{j}$ is a particle mass, $g_{j}$ is a
number of internal degrees of freedom):
         \eq{ p_{j}(T)~ = ~\frac{g_{j}}{6 \pi^2}
         \int_{0}^{\infty} k^{2}d k ~\frac {k^{2}}{\sqrt{k^2 +
 m_{j} ^2}}~\left[\exp{\left( \frac{\sqrt{k^2 + m_{j}^2}}{T} \right)} \pm 1\right]^{-1}
    \label{pressure}~,}
 where $+ 1$ is used for fermions and $-1$ for bosons.
The total pressure $p(T)$ is a sum of partial pressures $p_{j}(T)$
 over all particle species `$j$'. The energy density $\varepsilon(T)$ and
 entropy density $s(T)$ for the system with zero chemical
 potentials are calculated from the thermodynamical identities:
 $\varepsilon=Tdp/dT-p$ and $s=dp/dT$.  For massless particles one obtains:
$p(T) = \sigma/3 T^4~,~\varepsilon(T) =\sigma T^4~,~ s(T) = 4\sigma/3
T^{3}~,$ where $\sigma = (g^b + 7 g^f/8)\cdot\pi^{2}/30$ with $g^b$ and
$g^f$ being respectively the total number of degrees of freedom for bosons
and fermions. In the quark gluon plasma (QGP) we have $g_{Q}^{b}=2\cdot 8
=16$ and $g_{Q}^{f}= 2\cdot 2\cdot 2 \cdot 3 =24$ massless degrees of
freedom which correspond to the gluons and non-strange (anti)quarks,
respectively. The pressure of strange quarks and anti-quarks is given by
Eq.~(\ref{pressure}) with the degeneracy factor $g_{Q}^{s}=2\cdot 2\cdot
3= 12$ and quark mass $m_{s}=175$~MeV. We use the bag--model EoS for the
QGP \cite{bag}: $p_{Q}=p_{id} - B$, $\varepsilon_{Q}=\varepsilon_{id}+B$,
i.e. the constant term $B>0$ is subtracted from the total ideal gas
pressure and is added to the total ideal gas energy density.

In the confined (hadron) phase we assume the effective degeneracy factor
$g_{H}^{ns}$ for massless non-strange degrees of freedom and $g_{H}^{s}$
for strange degrees of freedom (the strange effective degrees of freedom
are assumed to be massive with mass $m_{H}$ close to the kaon mass). These
parameters of the confined phase are obtained in
Ref.~\cite{Gazdzicki:1998vd} from fitting the data on multiplicities of
strange and non-strange hadrons in Au+Au collisions at the AGS energies.

The temperature of the phase transition, $T_{c}$, is defined by the Gibbs
criterion of equal pressures: $p_{H}(T_{c})=p_{Q}(T_{c})$. At $ T = T_c $
the system stays in the mixed phase and its energy density equals to:
 \eq{ \varepsilon ~ =~(1-\xi)\cdot \varepsilon_H(T_{c})~ +~ \xi\cdot
 \varepsilon_Q(T_{c}) ~,\label{epsc}}
 where $0\le \xi \le 1$ is the part of the system occupied by the Q-phase.

 For the total strange particle number density in the $H$, $Q$ and mixed phases we
 have, respectively:
 \begin{eqnarray}\label{nsh}
 n^{s}_{H}(T)~& =&~\frac{g_{H}^{s}}{2\pi^{2}}~\int_{0}^{\infty}k^{2}d k~
 \exp\left(- ~ \frac{\sqrt{k^{2}+m_{H}^{2}}}{T}~\right)~=~
 \frac{g^{s}_{H}}{2 \pi^2}\cdot m_{H}^2 T\cdot
 K_{2}\left(\frac{m_{H}}{T}\right)~,\\
 n^{s}_{Q}(T) ~&=&~ \frac{g^{s}_{Q}}{2 \pi^2}~\int_{0}^{\infty}k^{2}d k~
 \left[\exp\left(\frac{\sqrt{k^{2}+m_{s}^{2}}}{T}\right)~+~1\right]^{-1}~,
 \label{nsq}\\
n^s_{mix}(\varepsilon)~ &=&~ (1-\xi)\cdot n_{H}^{s}(T_{c}) + \xi \cdot
n_{Q}^{s}(T_{c})~. \label{nsm}
\end{eqnarray}

\mysection{Statistical and Dynamical Fluctuations}

Various values of the energy $E$ and volume $V$ of the initial equilibrium
state lead to different, but uniquely determined, initial entropies $S$.
When the collision energy is fixed, the energy, which is used for particle
production still fluctuates. These fluctuations of the inelastic energy
are caused by the fluctuations in the dynamical process which leads to the
particle production. They are called the {\it dynamical} energy
fluctuations \cite{Gazdzicki:2003bb}. Clearly the {\it dynamical} energy
fluctuations lead to the {\it dynamical} fluctuations of any macroscopic
parameter of the matter.
%
%
In Ref.~\cite{Gazdzicki:2003bb} the dynamical {\it entropy} fluctuations
were considered and related to the dynamical energy fluctuations
and EoS as:
\begin{equation}\label{T}
R_{e}~\equiv ~\frac{(\delta S)^2/S^2}{(\delta E)^2/E^2} ~=~\left(1 +
\frac{p}{\varepsilon}\right)^{-2}~,
\end{equation}
providing the volume fluctuations were absent $\delta V= 0$.
%
The ratio $p/\varepsilon$ reaches a minimum -- the so called `softest
point' \cite{Hung:1994eq} of the EoS -- at the boundary between the mixed
phase and the QGP. Thus, we expect a non-monotonic energy dependence of
$R_{e}$ with a maximum at this boundary point. The numerical results for
the ratio $R_{e}$ (\ref{T}) calculated within the SMES in
Ref.~\cite{Gazdzicki:2003bb} are shown in Fig.~1.

The early stage energy and entropy
are not directly measurable. In an experiment only momenta and energies of
final state particles in a limited acceptance are reconstructed. Thus, the
question arises whether the early stage fluctuations can be reconstructed
from the experimentally accessible information. In
Ref.~\cite{Gazdzicki:2003bb} we argued that this may be possible in the
case of entropy, provide that the distortions caused by final state
correlations and statistical noise are removed. The distortions can be
minimized by a proper choice of the studied observables. For instance in
the case of study of dynamical entropy fluctuations it is better to
extract them from the fluctuations of negatively charged hadrons than from
all charged hadrons, as the latter is influenced by correlations due to
resonance decays and global charge conservation. 
%
As a method to extract the dynamical fluctuations from the statistical
noise the so-called sub-event method \cite{Voloshin:1999yf} can be used.
In this method the fluctuations are measured in two different
non-overlapping but dynamically equivalent regions of the phase space (see
Ref.~\cite{Gazdzicki:2003bb} for further details).

The idea to study the EoS of the matter created at the early stage
of collisions by an analysis of the dynamical fluctuations, proposed
originally for entropy \cite{Gazdzicki:2003bb}, can be extended  for
strangeness.
This subject
is discussed in the next section. 
We note that among different macroscopic 
parameters of the system 
the entropy and strangeness content are
of special importance \cite{Gazdzicki:1998vd}: they  are created at
the early stage of A+A collisions, they are sensitive to the EoS of the
matter and they are approximately conserved during the system
expansion, hadronisation and freeze-out.

\mysection{Fluctuations of Strangeness}

In this central section of the paper we study within the SMES the energy
dependence of the dynamical {\it strangeness} fluctuations caused by the
dynamical energy fluctuations. We define strangeness $N_{s}$ as a total
number of all strange and anti-strange particles and consider the
fluctuation ratio defined as:
\begin{equation}\label{rs}
R_{s}~=~\frac{(\delta \overline{N}_{s})^2/\overline{N}_{s}^2}{(\delta
E)^2/E^2}~.
\end{equation}
As in the case of entropy fluctuations we assume that experimental
procedure allows to eliminate the event--by--event fluctuations of the
initial volume $V$. Since $\overline{N}_{s}=n^{s}\cdot V$,
the dynamical
fluctuations of $\overline{N}_{s}$ are defined by that of the strangeness
density: $\delta\overline{N}_{s}/\delta E=\delta n^{s}/\delta
\varepsilon$ and consequently
$R_{s}~=~(\delta n^{s}/\delta \varepsilon)^2 \cdot (\varepsilon / n^{s})^2$.

The strangeness density depends on $T$ in the pure confined
and deconfined phases according to Eqs.~(\ref{nsh}) and (\ref{nsq}),
respectively. One has therefore to calculate the fluctuations of $n^{s}$
due to the fluctuations of temperature and then the fluctuations of $T$
due to those of $\varepsilon$:
 \eq{ \frac{\delta n^{s}_{(H,Q)}} {\delta\varepsilon}~ =~
 \frac{d n^{s}_{(H,Q)}}{d T}\cdot
 \frac{d T}{d \varepsilon}~.\label{nshq}}
In the mixed phase region, $\varepsilon_{H}(T_{c})<\varepsilon
<\varepsilon_{Q}(T_{c})$, the strangeness dynamical fluctuations can be
found from Eqs.~(\ref{epsc}) and (\ref{nsm}):
 \eq{\frac{\delta n^{s}_{mix}}{\delta \varepsilon}~ =~
  \frac{n^{s}_{Q}(T_{c}) - n^{s}_{H}(T_{c})}{\varepsilon_{Q}(T_{c})-\varepsilon_{H}(T_{c})}
 ~.\label{nsmm}}
We note that the fluctuation ratio (\ref{nsmm}) does not depend on the
energy density $\varepsilon$.

 Let us first find the asymptotic behavior of $R_{s}$ at high and small $T$.
When $T \rightarrow \infty$ the system is in the QGP phase. The strange
(anti)quarks can be considered as massless and the bag constant can be
neglected. Then $\varepsilon \propto T^4$ and $ n^s \propto T^3 $ and
consequently $d\varepsilon/\varepsilon=4 \cdot dT/T$ and $dn^{s}/n^{s}=3
\cdot dT/T$, which result in $R_{s}=(3/4)^{2}\cong 0.56$. In the confined
phase, $T<T_{c}$, the energy density is still approximately given by
$\varepsilon_{H}\propto T^{4}$ due to the dominant contributions of
non-strange hadron constituents. However, the dependence of strangeness
density on $T$ is dominated in this case by the exponential factor,
$n_{H}^{s}\propto \exp(-m_{H}/T)$, as $T<<m_{H}$. Therefore, at small $T$
one finds $d\varepsilon_{H}/\varepsilon_{H}\propto 4 \cdot dT/T$ and
$dn_{H}^{s}/n_{H}^{s}\propto m_{H} \cdot dT/T^{2}$, so that the ratio
$R_{s}\cong m_{H}/(4T)$ decreases with increasing temperature (energy
density). The strangeness density $n_{H}^{s}$ is small and goes to zero at
$T\rightarrow 0$, but the fluctuation ratio $R_{s}$ (\ref{rs}) is large
and goes to infinity at zero temperature limit.

The results of numerical calculations for the collision energy dependence
of $R_{s}$ (\ref{rs}) within SMES are presented in Fig.~2. As expected the
fluctuation ratio $R_{s}$ is independent of energy and equals
approximately 0.56 at high collision energies when the initial state
corresponds to the hot QGP. $R_{s}$ rapidly decreases at small collision
energies. A pronounced minimum-structure is observed in the dependence of
$R_{s}$ on the collision energy. It is located in the collision energy
region $30\div 60$~A$\cdot$GeV, where the mixed phase is created at the
early stage of A+A collision.

The ratio $R_{s}$ (\ref{rs}) is defined above for the infinitesimal energy
fluctuations. From Fig.~2 one observes the `discontinuities' of $R_{s}$ at
the boundaries of the pure hadron and QGP phases with the mixed phase. It
is easy to understand this result, as we have used different equations for
$n^{s}$ and $\varepsilon$ in each separate phase: the function $n^{s}$
does not have discontinuities, but its derivative $\delta n^{s}/\delta
\varepsilon$ does. In fact, the dynamical energy fluctuations are not
infinitesimal. Even more, the dynamical fluctuations should be not too
small to separate them safely from the statistical fluctuations. The
quantitative analysis presented below demonstrates that by introducing the
finite size of the energy fluctuations we avoid the artificial
discontinuities in the $R_{s}$ ratio.

We denote by $E$ an {\it average} energy of the created matter at fixed
collision energy. We further assume that event-by-event values of energy
$E^{\prime}$ vary around $E$ according to the Gauss distribution: \eq{
P(E^{\prime},E)~=~C~\exp\left[~-~\frac{(E-E^{\prime})^2}{2\sigma^2}\right]~,
\label{gauss}}
where $\sigma=a E$ with $a=const$, and $C\cong (2\pi\sigma^{2})^{-1/2}$ is
defined by the normalization condition
 $\int _0^{\infty} dE^{\prime}P(E^{\prime},E)=1.$
 The dynamical averaging of any observable $f(E^{\prime})$
can be defined now as:
\eq{\langle f\rangle~=~\int_0^{\infty}
dE^{\prime}~P(E^{\prime},E)~f(E^{\prime})~. \label{av}}
The ratio $R_{s}$ for the finite energy fluctuations (\ref{gauss}) is
given by
 \eq{ R_{s} ~\equiv~\frac{\langle(\delta N_s)^2\rangle /\langle
N_s\rangle ^2} {\langle(\delta E)^2\rangle/\langle E \rangle ^2}~,
\label{rsav}} where $(\delta X)^{2}\equiv (X-\langle X \rangle)^{2}$.
Providing $E >> \sigma$ one gets $ \langle E \rangle \equiv
\int_0^{\infty}dE^{\prime}P(E^{\prime},E)~E^{\prime} \cong E$ and
$\langle(\delta E)^2 \rangle = \langle E^2\rangle - \langle E\rangle^2
\cong (\sigma)^{2}$, and consequently $\langle(\delta E)^2\rangle /
\langle E \rangle ^2 = (\sigma / E)^2 = a^{2}$, i.e. for distribution
(\ref{gauss}) the relative fluctuations are independent of the collision
energy.

The function $N_{s}(E^{\prime})$ needed to calculate $\langle
N_{s}\rangle$ and $\langle N_{s}^{2}\rangle$ is defined by
Eqs.~(\ref{nsh}-\ref{nsm}) with $\varepsilon \equiv E^{\prime}/V$. The
results of numerical calculations of $R_{s}$ (\ref{rsav}) with $\sigma /
E= a = 0.1$ are shown in Fig.~2. The qualitative features of the $R_{s}$
dependence on the collision energy are not changed by the introduction of
finite energy fluctuations (very similar plots are obtained for $a=0.05$
and $a=0.2$). The main difference is the disappearance of the sharp edges
characteristic for infinitesimal fluctuations.

We apply now our procedure of averaging over finite energy fluctuations to
the ratio $R_{s}$ (\ref{T}). The calculations are done according to
Eq.~(\ref{av}) with $S(E^{\prime})=V (\varepsilon + p)/T$:
\begin{equation}
\langle S^{n} \rangle=\int_0^{\infty}dE^{\prime}
~P(E^{\prime},E)~S^{n}(E^{\prime})~,~~n=1,2;~~~~ \langle (\delta
S)^2\rangle ~ =~ \langle S^2\rangle - \langle S \rangle ^2~.
\end{equation}
The dependence of $R_e$ on the collision energy for distribution
(\ref{gauss}) with $\sigma / E = a=0.1$ is shown in Fig. 1 by dashed line.
Again, as in the case of $R_s$, the averaging procedure does not change
the basic features of the dependence.

Both the entropy and strangeness fluctuation measures, $R_e$ and $R_s$,
show 'anomalous' behavior in the transition region: the maximum is
observed for $R_e$ and the minimum for $R_s$. Consequently, even stronger
anomaly is observed in the ratio: 
\eq{ R_{s/e}~\equiv ~\frac{ R_{s}}{
R_e}~ =~\frac{\langle (\delta N_s)^2\rangle /\langle N_s\rangle ^2}
{\langle (\delta S)^2\rangle /\langle S\rangle ^2}~ \cong ~\frac{\langle
(\delta N_{s})^{2}\rangle /\langle N_{s}\rangle ^{2}}{\langle (\delta
N_{-})^{2}\rangle /\langle N_{-}\rangle ^{2}}~.\label{R}}
Its dependence on the collision energy is shown in Fig. 3 for infinitely
small energy fluctuations and for $\sigma / E = a=0.1$. Experimental
measurements of $ R_{s/e} $ may be easier than the measurements of $R_e$
or $R_s$ because the ratio $ R_{s/e} $ requires measurements of particle
multiplicities only, whereas both $R_e$ and $R_s$ involve also
measurements of particle energies.
Finally we note that the relative dynamical fluctuations of
strangeness and entropy are approximatelly equal 
(see \cite{Gazdzicki:2003bb}) to
the corresponding fluctuations of $K^+$ meson and negatively
charged hadron multiplicities, respectively. Thus the $ R_{s/e} $
fluctuation ratio can be expressed by the measurable quantities as:
\eq{ R_{s/e}
\cong ~\frac{\langle
(\delta N_{K^+})^{2}\rangle /\langle N_{K^+}\rangle ^{2}}{\langle (\delta
N_{-})^{2}\rangle /\langle N_{-}\rangle ^{2}}~.\label{RR}}

 \mysection{Conclusions}

We have considered strangeness fluctuations as a potential probe of the
equation of state and microscopic content of the strongly interacting
matter created at the early stage of high energy nucleus--nucleus
collisions. In order to quantify the fluctuations we have introduced a new
measure $ R_{s/e} $ (\ref{R}) constructed from fluctuations of strange and
non--strange hadron multiplicities. We have predicted, within statistical
model of the early stage \cite{Gazdzicki:1998vd}, the energy dependence of
the $ R_{s/e} $ measure and we have shown that it is strongly sensitive to
the form of matter created at the early stage of nucleus--nucleus
collisions. In particular, a ``tooth'' structure (see Fig.~3) 
is expected in the
collision energy domain $30\div 60$~A$\cdot$GeV in which deconfinement
phase transition is located.

\section*{Acknowledgments.}
Partial support by Bundesministerium f\"ur Bildung und Forschung (M.~G.)
and by Polish Committee of Scientific Research under grant 2P03B04123
(M.~G.) 
and by  DAAD scholarship under
 Leonhard-Euler-Stipendienprogramm (O.S.Z.)
is acknowledged.

%
%
%
%
%
%
%
%
%
%
%

\begin{figure}
\hspace{1cm}
\epsfig{file=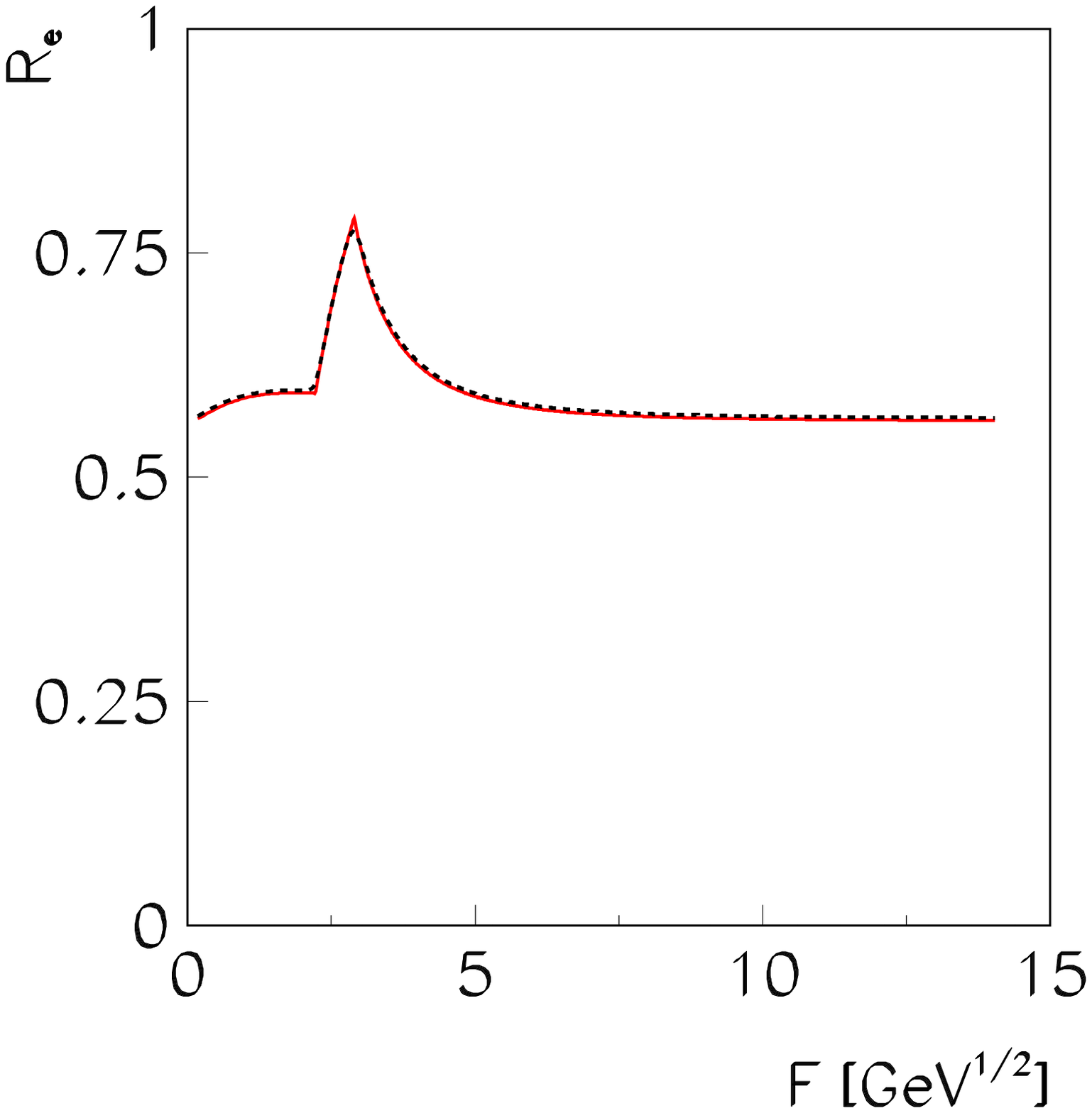,width=130mm}
\vspace{1cm}
\caption{
The dependence of the entropy to energy fluctuations, $R_e$,
calculated within SMES \protect\cite{Gazdzicki:1998vd} on Fermi's
collision energy measure $F$ ($F \equiv (\sqrt{s_{NN}} - 2
m_N)^{3/4}/\sqrt{s_{NN}}^{1/4}$, where $\sqrt{s_{NN}}$ is the c.m.s.
energy per nucleon--nucleon pair and $m_N$ the rest mass of the nucleon).
The non--monotonic behavior, the ``shark fin'' structure, is caused by the
modification of fluctuations
expected in the vicinity of the mixed phase region.
The solid and dashed lines indicate results for infinitely small and
finite ($\sigma / E = 0.1$) energy fluctuations, respectively.
}
\label{entropy}
\end{figure}

\newpage

\begin{figure}
\hspace{1cm}
\epsfig{file=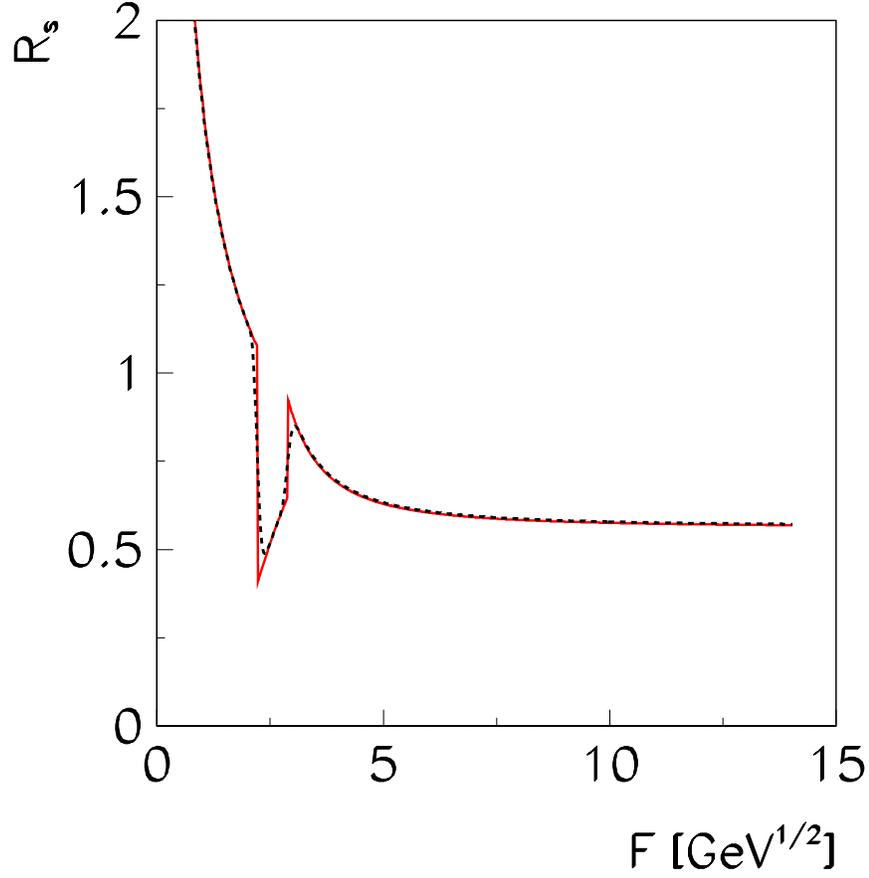,width=130mm}
\vspace{1cm}
\caption{
The dependence of the strangeness to energy fluctuations, $R_s$,
calculated within SMES \protect\cite{Gazdzicki:1998vd} on Fermi's
collision energy measure $F$ ($F \equiv (\sqrt{s_{NN}} - 2
m_N)^{3/4}/\sqrt{s_{NN}}^{1/4}$, where $\sqrt{s_{NN}}$ is the c.m.s.
energy per nucleon--nucleon pair and $m_N$ the rest mass of the nucleon).
The non--monotonic behavior is caused by the modification of
fluctuations expected in the vicinity of the mixed phase region.
The solid and dashed lines indicate results for infinitely small and
finite ($\sigma / E = 0.1$) energy fluctuations, respectively.}
\label{str}
\end{figure}

\newpage

\begin{figure}
\hspace{1cm}
\epsfig{file=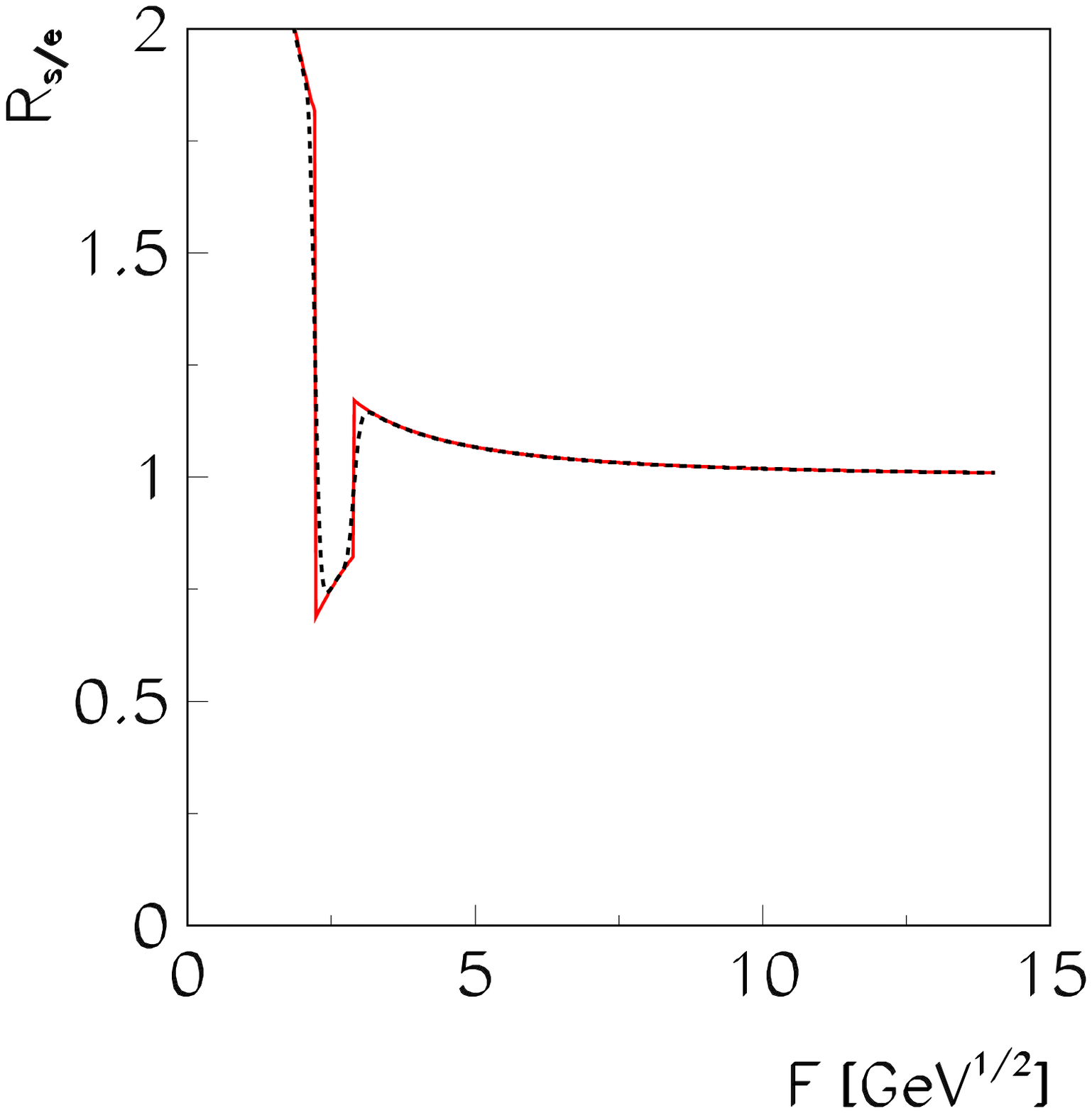,width=130mm}
\vspace{1cm}
\caption{
The dependence of the strangeness to entropy fluctuations, $R_{s/e}$,
calculated within SMES \protect\cite{Gazdzicki:1998vd} on Fermi's
collision energy measure $F$ ($F \equiv (\sqrt{s_{NN}} - 2
m_N)^{3/4}/\sqrt{s_{NN}}^{1/4}$, where $\sqrt{s_{NN}}$ is the c.m.s.
energy per nucleon--nucleon pair and $m_N$ the rest mass of the nucleon).
The non--monotonic behavior, the ``tooth'' structure, is caused by the
modification of
fluctuations expected in the vicinity of the mixed phase region.
The solid and dashed lines indicate results for infinitely small and
finite ($\sigma / E = 0.1$) energy fluctuations, respectively.}
\label{se}
\end{figure}
\end{document}